\documentclass{iau_FM}
\usepackage{graphicx,wrapfig}

\newcommand\ltsim{\raisebox{-.5ex}{$\;\stackrel{<}{\sim}\;$}}
\newcommand\gtsim{\raisebox{-.5ex}{$\;\stackrel{>}{\sim}\;$}}

\title[FM 10.~~Stripped-envelope supernova rates] %% give here short title %%
{Stripped-envelope supernova rates \\ and host-galaxy properties}

\author[Graur et al.]   %% give here short author list %%
{Or Graur$^{1,2}$, F.B. Bianco$^1$, M. Modjaz$^1$, D. Maoz$^3$, I. Shivvers$^4$, A.V. Filippenko$^4$, and W. Li$^4$}

\affiliation{$^1$CCPP, New York University, 4 Washington Place, New York, NY 10003, USA \\ email: {\tt orgraur@nyu.edu} \\ [\affilskip]
             $^2$Dept. of Astrophysics, American Museum of Natural History, New York, NY 10024, USA \\ [\affilskip]
             $^3$School of Physics and Astronomy, Tel-Aviv University, Tel-Aviv 69978, Israel \\ [\affilskip]
             $^4$Department of Astronomy, University of California, Berkeley, CA 94720-3411, USA}
             
\pubyear{2015}
\jname{Astronomy in Focus, Volume 1} 
\editors{Piero Benvenuti, ed.}

\begin{document}

\maketitle

\begin{abstract}
The progenitors of stripped-envelope supernovae (SNe Ibc) remain to be conclsuively identified, but correlations between SN rates and host-galaxy properties can constrain progenitor models. Here, we present one result from a re-analysis of the rates from the Lick Observatory Supernova Search. Galaxies with stellar masses $\ltsim10^{10}~{\rm M_\odot}$ are less efficient at producing SNe Ibc than more massive galaxies. Any progenitor scenario must seek to explain this new observation. 
\end{abstract}

\begin{figure*}
 \centering
 \includegraphics[width=0.8\textwidth]{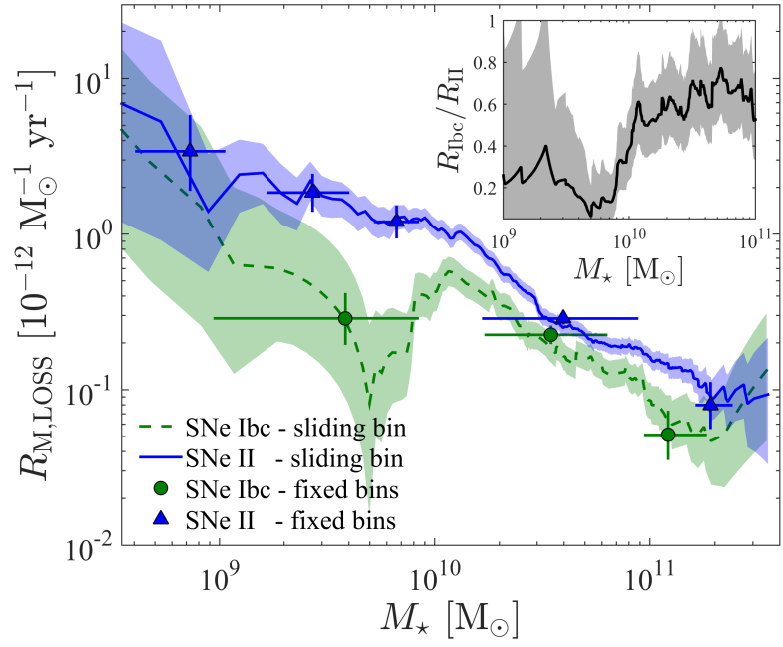}
  \caption{Mass-normalized CC SN rates versus galaxy stellar mass, measured in sliding (curves) and fixed (symbols) bins. The shaded patches are the 68\% Poisson uncertainties of the rates as measured with the sliding bin. The inset shows the ratio between the SN Ibc and SN II rates. In galaxies less massive than $\sim10^{10}~{\rm M_\odot}$, the SN Ibc rates plateau and are deficient relative to the SN II rates.}
  \label{fig1}
\end{figure*}

Core-collapse supernovae (CC SNe) are broadly divided into SNe II and SNe Ibc (also known as stripped-envelope SNe; see, e.g., \cite[Modjaz et al. 2014]{2014AJ....147...99M}). The progenitors of the latter have yet to be unambiguously identified (\cite[Eldridge et al. 2013]{2013MNRAS.436..774E}). Knowledge of the progenitors will impact various field of astrophysics, from the explosion processes of these stars, through galaxy evolution and cosmic nucleosynthesis. In past works (e.g., \cite[Graur \& Maoz 2013, Graur et al. 2015]{GraurMaoz2013,2015MNRAS.450..905G}), we have shown that correlating SN rates with various host-galaxy properties can constrain progenitor scenarios, by, e.g., recovering the delay-time distribution of SNe Ia. A fraction of the Lick Observatory Supernova Search (LOSS; \cite[Leaman et al. 2011, Li et al. 2011a,b]{2011MNRAS.412.1419L,li2011LF,li2011rates}) SN host galaxies have Sloan Digital Sky Survey spectra, from which various galaxy properties were extracted. We have used this fraction of LOSS SNe to re-measure CC SN rates and test for correlations with different galaxy properties. In Fig.\,\ref{fig1}, we show CC SN rates as a function of galaxy stellar mass, $M_\star$, from which it appears that while in galaxies with $M_\star\gtsim10^{10}~{\rm M_\odot}$ the rates of SNe Ibc track those of SNe II, in less-massive galaxies the SN Ibc rates are deficient relative to the SN II rates. This deficiency hints that some physical proccess(es) related to the progenitors of SNe Ibc is hindering their production in galaxies less massive than $\sim10^{10}~{\rm M_\odot}$. Any SN Ibc progenitor model must reproduce these rates and explain this deficiency.

\end{document}